\documentclass[showpacs,prb,twocolumn,superscriptaddress,floatfix,amsmath]{revtex4}
\usepackage{graphicx,amsmath}
\begin{document}
\title{ 
{\bf Monte Carlo simulations of two-dimensional charged 
bosons}}
\author{S. De Palo}
\affiliation{ Dipartimento di Fisica, 
Universit\`a di Roma ``La Sapienza'', P.le Aldo Moro 5, 00185 Rome, Italy}
\author{S. Conti}
\affiliation{Max Planck Institute for Mathematics in the Sciences,
Inselstrasse 22, 04103 Leipzig, Germany}
\author{S. Moroni}
\affiliation{SMC INFM, Dipartimento di Fisica,
Universit\`a di Roma ``La Sapienza'', P.le Aldo Moro 5, 00185 Rome, Italy}
\date{\today}
\begin{abstract}
Quantum Monte Carlo methods are used to calculate various 
ground state properties of charged bosons in two dimensions,
throughout the whole density range where the fluid phase is stable.
Wigner crystallization is predicted at $r_s\simeq 60$. 
Results for the ground state energy and the momentum distribution
are summarized in analytic interpolation formulas embodying known asymptotic 
behaviors. Near freezing, the condensate fraction is less than 1\%. 
The static structure factor $S(k)$ and susceptibility
$\chi(k)$ are obtained from the density-density correlation 
function in imaginary time, $F({\bf k},\tau)$.
An estimate of the energy of elementary excitations, given
in terms of an upper bound involving $S(k)$ and $\chi(k)$, is
compared with the result obtained via analytic continuation 
from $F({\bf k},\tau)$.   
\end{abstract} 
\maketitle
\section{Introduction}

The two-dimensional fluid of point-like spinless bosons interacting with
a $1/r$ potential has drawn attention in the literature\cite{saarela}
as a model in quantum statistical mechanics which parallels the physically 
more relevant fluid of electrons.
At zero temperature, the model is specified by the coupling parameter
$r_s=1/{\sqrt{\pi n}a_B}$, where $n$ is the density and $a_B$ the Bohr
radius. For small $r_s$ the system is a weakly coupled fluid, well 
described by the Random Phase Approximation,\cite{hines}
whereas it becomes strongly correlated and eventually undergoes Wigner 
crystallization upon increasing $r_s$.
Several results for the ground state energy, static structure, screening
properties and elementary excitations have been reported using the
Correlated Basis Function theory,\cite{sim,saarela} various implementation
of the Singwi-Tosi-Land-Sj\"olander (STLS) formalism,\cite{um,gold} and
the Overhauser model.\cite{asgari} The momentum distribution has 
been calculated for low $r_s$ in the Bogolubov approximation.
\cite{strepparola}
A comparison between the STLS results for the $1/r$ potential 
and the ln$(r)$ potential has been reported by Moudgil {\it et al}.
\cite{Mougdil}

Although the charged boson model may find applications to
superconductors, either as a system of bound electron pairs\cite{polarons}
or in terms of an effective action with Fermionic degrees of freedom
integrated out,\cite{stefania} no direct realization of the system is
experimentally available. Therefore numerical results provided by
quantum Monte Carlo (QMC) simulations constitute the only reliable benchmark for
analytic approaches. 
Extensive simulation results are available for 3D charged 
bosons\cite{alder,conti} and for the 2D system with the 
ln$(r)$ interaction.\cite{magro1,magro2,blatter}

In this work we present QMC results for several ground state properties
of the 2D fluid of charged bosons with the $1/r$ potential. 
We use two different algorithms, namely diffusion Monte Carlo 
(DMC),\cite{mitas}
which is more efficient in the calculation of mixed averages,
and reptation quantum Monte Carlo (RQMC),\cite{baroni} which gives easier
access to correlations in imaginary time. 
The exact ground state energy and the mixed estimate\cite{mitas} 
of the one-body density matrix are calculated with the former.
Unbiased estimates of the static structure factor and the susceptibility are 
instead obtained, using RQMC, from the auto-correlation in imaginary
time of the density fluctuation operator.
The inverse Laplace transform of the same auto-correlation function
yields valuable information on the spectrum of elementary excitations. 

\section{Method}

Quantum Monte Carlo is
the method of choice for strongly interacting bosonic systems
in their ground state, because it yields exact numerical results 
for a number of quantities, subject only to known statistical errors. 

The DMC method\cite{mitas} samples a probability distribution
proportional to the ``mixed distribution'' $f(R)=\Phi(R)\Psi(R)$, where 
$R=\{{\bf r}_1,\cdots,{\bf r}_N\}$ is a point in the $2N$-dimensional
configuration space of the system,
$\Psi(R)$ is a trial wave-function, and 
$\Phi(R)$ is the ground-state wave-function.
The exact ground state energy is obtained as the average over the mixed
distribution of the local energy, $E_L(R)=\Psi(R)^{-1}H\Psi(R)$.
For a general operator not commuting with the Hamiltonian, 
ground-state averages can be approximated by the extrapolated estimate
(twice the average over the mixed distribution minus the variational
estimate),\cite{mitas} which leads to an error quadratic in the 
difference $(\Phi-\Psi)$.
Our results for the one-body density matrix are given in terms of this
extrapolated estimate, as in Ref.~\onlinecite{conti}.

For operators diagonal in $R$ we avoid mixed estimates
resorting to the RQMC method\cite{baroni} (one could alternatively use
the forward walking technique\cite{fw} within the DMC method).
In RQMC, the evolution in imaginary time of the system is represented
by a time-discretized path $X=$\{$R_0, \cdots, R_M$\}.
The algorithm samples the distribution $P(X)=\Psi(R_0)^2\Pi_{i=1}^MG(R_{i-1}\to
R_i;\epsilon)$, where $G(R\to R'; \epsilon)$ is a short-time approximation
to the importance-sampled Green's function
$\Psi(R')\langle R'|\exp(-\epsilon H)|R\rangle \Psi(R)^{-1}$.
Assuming $M$ is large enough, the inner time slices of the path
are individually sampled from the distribution $\Phi(R)^2$, and
sequentially sampled according to the quantum dynamical fluctuations
in the ground state. Pure estimators, 
$\langle\Phi|O|\Phi\rangle=\langle\langle O(R_i)\rangle\rangle$,
and imaginary-time correlation functions, 
$c(\tau)=\langle\langle O(R_i)O(R_{i+n})\rangle\rangle$,
are thus readily accessible (here $\langle\langle\cdot\rangle\rangle$
means average over the random walk in the space of quantum paths $X$,
and $\tau=n\epsilon$).\cite{baroni}

In all simulations we consider a system of $N$ particles in a square cell with periodic 
boundary conditions. The trial function is chosen of the pair 
product form, $\Psi(R)=\exp(-\sum_{ij}(u|{\bf r}_i-{\bf r}_j|))$, where 
$u(r)$ is the RPA pseudopotential following Ref.~\onlinecite{Cep_78}. 
Both the pseudopotential and the Coulomb interaction are evaluated using 
generalized Ewald sums.\cite{Cep_78} As usual,\cite{alder,Cep_78} we 
estimate the finite-size effect 
on the ground-state energy from variational Monte Carlo simulations.
Variational energies $E_N$, calculated with 
$N$ in the range $25$--$200$, are used to determine the best--fit 
parameter in the form $E_{\infty} = E_N + a(r_s)/N+b(r_s)/N^2$. 
Assuming that the same size dependence holds for the exact DMC energies,
the optimal parameters $a(r_s)$ and $b(r_s)$ are then used to extrapolate
to the thermodynamic limit the result of a single DMC simulation with $N=52$.
Other quantities have comparatively smaller finite-size errors, typically 
below the statistical accuracy of the present simulations.

\section{Results}

\subsection{Ground-state energy}\label{ene}

The DMC ground state energies of the 
2D bosonic fluid in the thermodynamic limit are compared in 
Table \ref{tab_ene} with the results obtained with the 
Singwi--Tosi--Land--Sj\"olander (STLS) method by Gold,\cite{gold} with a 
parametrized wave function approach by Sim, Tao and Wu\cite{sim} and 
within the Hypernetted Chain Approximation (HNC) by Apaja {\it et 
al.}.\cite{saarela} While all computations agree qualitatively, we 
note that the agreement between HCN and the exact DMC results is 
particularly good. 
Our DMC results can be 
accurately reproduced by the parametrized function: 
\begin{equation}E_g(r_s)=-[a_0 r_s^{b_0}+ a_1 r_s^{b_1}+ 
a_2 r_s^{b_2}+ a_3 
r_s^{b_3}]^{-c}\label{eqbos2drs0}\end{equation}where 
$a_0$ and $b_0$ are fixed by the small $r_s$ behavior \cite{saarela} 
($E(r_s\rightarrow 0) \simeq -1.29355/r_s^{2/3}$), $b_1$ 
is fixed requiring a constant sub-leading term for $r_s \rightarrow 
0$, $b_2$ and $b_3$ by requiring leading terms in $r_s^{-1}$ and $r_s^{-3/2}$ 
for $r_s \rightarrow \infty $. The final values of the parameters are
$c=7/40$, $a_0=0.2297$, $a_1=0.161$, $a_2=0.0594$, 
$a_3=0.01017$, $b_0=80/21$, $b_1=94/21$, $b_2=73/14$ and $b_3=40/7$. 
The reduced $\chi^2$ for the fit with 4 parameters and 7 data points
is $1.5$ at $r_s=1$.
The above interpolation formula allows to 
obtain, by means of the virial theorem, the unbiased estimator of the 
average kinetic energy $\langle ke\rangle =-d(r_s E_g)/dr_s$ as 
well as of the inverse compressibility $1/\rho K_T=-\frac{r_s}{4}
[\frac{\partial E_g}{\partial r_s}-r_s\frac{\partial^2 E_g}{\partial r_s^2}]$, 
both reported in Table \ref{tab_ene}. 
\begin{figure}
\includegraphics[width=67mm]{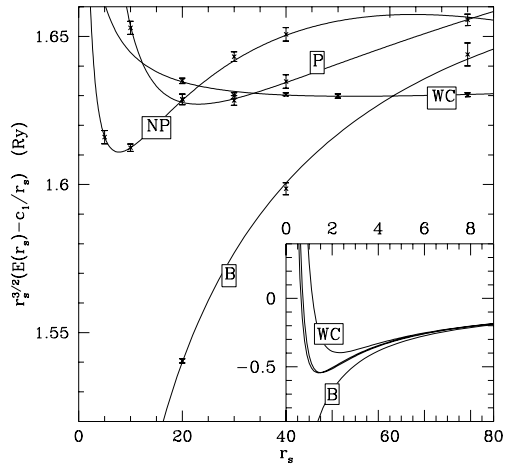}
\caption{
Ground-state energy for 2D triangular Wigner crystal (WC), 
bosons (B), unpolarized (UP) 
and polarized (P) fermions as a function of $r_s$. Wigner crystal and 
fermion data are from Ref.~\onlinecite{Rapi}. On purpose of clarity we 
plotted $r_s^{3/2}(E(r_s)-c_1/r_s))$, with $c_1=-2.2122$, while 
the inset shows the corresponding $E(r_s)$ curves. Points with error bars 
are size-extrapolated DMC results, continuous curves are analytical fits. 
}
\label{all_ene}
\end{figure}

In Fig.~(\ref{all_ene}) our results are  
compared with the previous DMC results by Rapisarda and 
Senatore \cite{Rapi} for 2D fermions and for the 2D Wigner crystal.
In two dimensions bosons 
crystallize at $r_s\simeq 60$ and fermions at $r_s\simeq 34$. 
The difference in critical density is 
analogous to the difference obtained in the 3D 
case, where bosons crystallize at $r_s=160$ and fermions at 
$r_s=100$.\cite{alder}
\begin{table*}
\caption{Ground state energy for 
bosons from VMC and DMC, extrapolated to the bulk limit and compared with 
estimates from approximate theories. We also give the
average kinetic energy and inverse compressibility obtained from
Eq.~(\protect\ref{eqbos2drs0}). All values are in Rydberg 
per particle, the digits in parenthesis represent the error bar in 
the last digit.}
\begin{tabular}{rlllllllll}\hline\hline
$r_s$ &  &  $E^{(DMC)}$ &  $E^{(VMC)}$ &  HNC\cite{saarela}    &  STW\cite{sim}     
       &  STLS\cite{gold}        & $\langle ke\rangle$ &  1/n$K_T$ 
\\ \hline 
1  &  & -1.1448(5)  & -1.14269(7)  & -1.1458  & -1.1062   & -        & 0.2903   & -0.531 \\
2  &  & -0.6740(2)  & -0.67192(6)  & -0.6740  & -0.6631   & -0.6484  & 0.1442   & -0.3582 \\
5  &  & -0.31903(5) & -0.317456(6) & -0.3185  & -0.3133   & -0.3078  & 0.04896  & -0.187  \\
10 &  & -0.17480(5) & -0.17385(3)  & -0.1741  & -0.16685  & -0.1724  & 0.01961  & -0.1097 \\
20 &  & -0.093387(8)& -0.092903(3) & -0.0928  & -0.086024 & -0.0959  & 0.007533 & -0.06177 \\
40 &  & -0.048986(8)& -0.048737(2) & --       & --        & --       & 0.00286  & -0.03359 \\
75 &  & -0.026965(6)& -0.0268246(8)& --       & --        & --       &
       0.001189 & -0.01892 \\ 
\hline\hline\label{tab_ene}\end{tabular}
\end{table*}
\subsection{Momentum distribution }   
The one--body density matrix $n(r)$ and its Fourier transform, 
the momentum distribution $n(k)$, have been 
computed performing random displacements of particles
on the sampled configurations as explained 
in Ref.~\onlinecite{magro1}.

\begin{figure}\includegraphics[width=67mm]{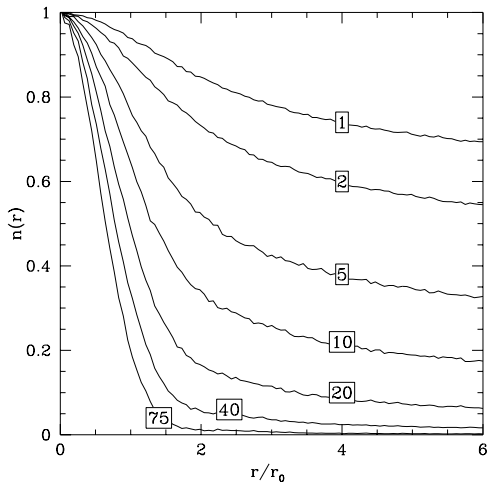}
\caption{One--body density matrix $n(r)$ at $r_s=$1, 2, 5, 10, 20, 40 
and 75 } 
\label{nr}
\end{figure}

At variance with the 3D case,\cite{conti} the standard procedure 
leads to strong size
effects due to the slow convergence of $n(r)$ to its asymptotic limit
$n_0=\lim_{r\to\infty} n(r)$. We removed the size-effect adopting the
correction proposed by Magro and Ceperley
\cite{magro1} for 2D bosons with $\ln r$ interactions. 
Our results for the one--body density matrix are shown in
Fig.~(\ref{nr}).

Extending to the 2D case the discussion presented for 3D charged bosons 
in Ref.~\onlinecite{cct96}, we fix the divergence of the momentum distribution
at small $k$
\begin{equation}
n(k\rightarrow 0) \simeq 
\frac{n_0}{4 S(k)}\simeq \frac{n_0\sqrt{r_s/2}}{(kr_0)^{3/2}}
\end{equation}
where $n_0$ is condensate fraction, and $r_0=r_s a_B$.
The cusp condition\cite{Kimball2D} instead 
gives information on the short--range behavior of the momentum
distribution:
\begin{equation}n(k\to\infty) \simeq {4 r_s^2 g(0)\over 
(kr_0)^6}\label{eqkimball2d}\end{equation} 
where $g(0)$ is the pair correlation function at $r=0$.
Moreover, at small density, we expect the momentum distribution to be
approximately Gaussian, in agreement with harmonic theory for the
crystalline phase.

We have collected all this information in a fitting  
formula to interpolate the DMC data for the momentum distribution $n(k)$:
\begin{eqnarray*}n(k) &=& (2\pi)^2\rho 
n_0\delta^2(k) +{n_0\sqrt{r_s/2}\over \kappa^{3/2}} e^{-\kappa^2/a_0^2}+ 
{4 g(0) r_s^2\over    a_1^6 + \kappa^6}\\&+& \left({a_2\over 
\sqrt {\kappa}} + a_3+ a_4\sqrt {\kappa}  + a_5 \kappa
\right)e^{-(\kappa^2-\kappa a_6)/a_7^2}\label{eq2bnkfit}\end{eqnarray*}
where $\kappa=kr_0$. 
Given the 
known values of the density and of $g(0)$ (see next section), we 
determined the remaining parameters by a least--squares fit to the DMC 
data on $n(k)$, $n(r)$ and on the average kinetic energy. 

Table \ref{nk_fit} contains the best--fit parameters 
and the resulting value of the condensate fraction $n_0$. 
\begin{table}
\caption{Best fit parameters for equation (\ref{eq2bnkfit}). 
The last line reports the value of $g(0)$ from 
Fig.~(\ref{gofr}) as used in the fit of 
$n(k)$.}
\begin{center}\begin{tabular}{lllllllll}\hline\hline
$r_s$ & &      1 & 2      & 5       & 10    & 20      & 40         & 75     \\ \hline
$ n_0 $& & 0.531  & 0.38  & 0.176 & 0.0677  & 0.018 & 0.001 & 0.0007 \\
$ a_0 $& & 0.839 & 0.853 & 0.475 & 0.977 & 0.861 & 1.21 & 2.59 \\
$ a_1 $& & 44     & 3.5   & 5.46  & -- & -- & --    & -- \\
$ a_2 $& & -0.086 & 0.492 & 2.17  & 1.96   & 1.05   & 0.946 & 0.098 \\
$ a_3 $& & 0.696  & 0.56  & -2.1  & -1.13   & -0.08 & -0.74 & 0.627 \\
$ a_4 $& & 1.13   & 0.226 & 0.23  & -0.01 & -0.163 & 0.184 & -0.103 \\
$ a_5 $& & 0.135  & 0.192 & 0.28  & 0.7   & -0.006 & -0.014 & -0.024 \\
$ a_6 $& & -111 & -6.07 & 1.12 & -1.52 & 0.849 & 3.44 & 0.576 \\
$ a_7 $& & 6.98 & 2.29 & 1.45 & 1.86 & 2.61 & 1.99 & 2.59 \\
$g_0$ & &  0.21   & 0.078 & 0.01  & -       & -      &       -    & -         \\ 
\hline\hline
$n_0$\cite{strepparola} & &  0.537  & 0.398  & 0.230    & -      & -
&       -    & -  \\  
\hline\hline
\label{nk_fit}\end{tabular}\end{center}\end{table}
The condensate fraction  decreases very rapidly with increasing $r_s$, the
depletion being already 50\% at $r_s=1$, in agreement with the result
of the Bogolubov theory\cite{strepparola} (in 3D\cite{conti} 
a similar depletion occurs at $r_s=5$). For large couplings, 
the Bogolubov theory
overestimates the condensate fraction. In a wide density range in the liquid
phase, say $r_s > 20$, $n_0$ is of the order of 1\% or less. Such small
values, obtained by fitting Eq.~\ref{eq2bnkfit}
to the extrapolated estimates from the simulation, are presumably
meaningful only as an indication of the order of magnitude.

\subsection 
{Imaginary-time correlation functions: static response function and 
static structure factor}

Information on charge response 
properties of the system like screening, plasma oscillations or 
polarization are contained in the imaginary time density-density 
correlation 
function:\begin{equation}F(k,\tau)=\frac{1}{N}\langle \rho_{{\bf 
k}}(\tau)\rho_{{\bf k}}(0)\rangle= 
\frac{1}{2\pi}\int^{\infty}_{-\infty} e^{-\tau\omega} 
S(k,\omega)\label{F_kt},\end{equation}
where $\rho_{{\bf 
k}}(\tau)=\sum_{i}e^{i({\bf k}\cdot{\bf r} )}$ and $S(k,\omega)$ is the 
dynamical response function. The correlation functions $F(k,\tau)$ have
been computed with RQMC for systems of 56 particles.

\begin{figure}[tb]
\includegraphics[width=67mm,angle=-90 
]{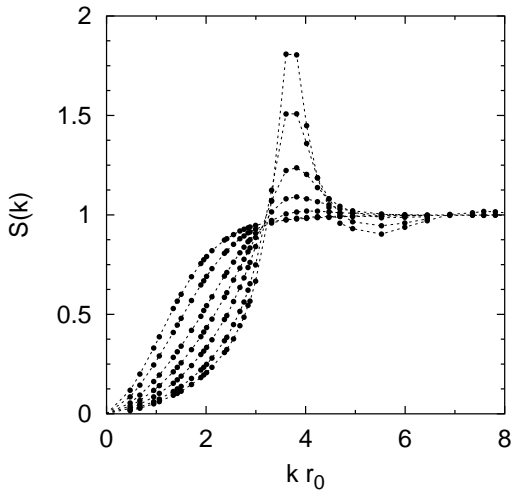}
\caption{Static structure factor $S(k)$ 
as a function of $kr_0$ for $r_s=1,2,5,10,20,40,60$.
Lines are only guide to the eyes. }\label{Fig3}\end{figure}

The static structure factor $S(k)$ is readily obtained from the 
imaginary-time density-density correlation function as:
\begin{equation}S(k)=\int_0^\infty d\omega~S(k,\omega)=F(k,0).
\end{equation}
In Fig.~(\ref{Fig3}) we report the 
behavior of $S(k)$ for various densities. As $r_s$ increases and 
approaches the crystallization density, a sharp peak develops in 
correspondence with the first lattice wave-vector of the 2D Wigner 
crystal, $kr_0= ( 2\pi\sqrt{3} )^{1/2}\simeq 3.3 $.

\begin{figure}[tb]
\includegraphics[width=67mm,angle=-90]{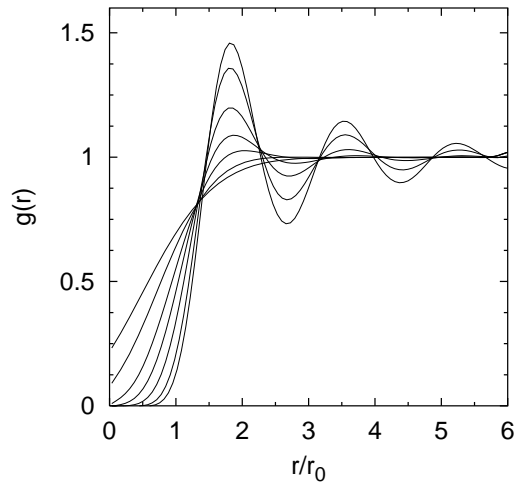}
\caption{The 
pair-distribution function for $r_s=1,2,5,10,20,40$ and $60$ (cubic spline 
interpolation of Monte Carlo data). Higher peaks correspond to higher 
values of $r_s$.
}\label{gofr}
\end{figure}

In Fig.~(\ref{gofr}) we report the pair distribution function: 
\begin{equation}g(r)=\frac{1}{N \rho} \sum_{i \ne j}\langle 
\delta(|{\bf 
r}_i-{\bf r}_j | -r)\rangle.\end{equation}
At low density $g(r)$ develops a high peak and long-range oscillations
typical of a system approaching localization.
As the density increases the effective 
repulsion between particles decreases and overlapping between charges 
becomes possible. 
The behavior of $S(k)$ and $g(r)$ 
is qualitatively in agreement with the findings of Apaja {\it et 
al.},\cite{saarela} but for both functions the Monte Carlo results 
show more pronounced effects of correlations at low densities.

The static
response function $\chi(k)$ can be evaluated from the relation 
\begin{eqnarray} \chi(k) =-2\int_0^\infty \frac{S(k,\omega)}{\omega} d\omega
=-2\int_0^\infty F(k,\tau) d\tau.
\end{eqnarray}

\begin{figure}[tb]
\includegraphics[width=67mm,angle=-90]{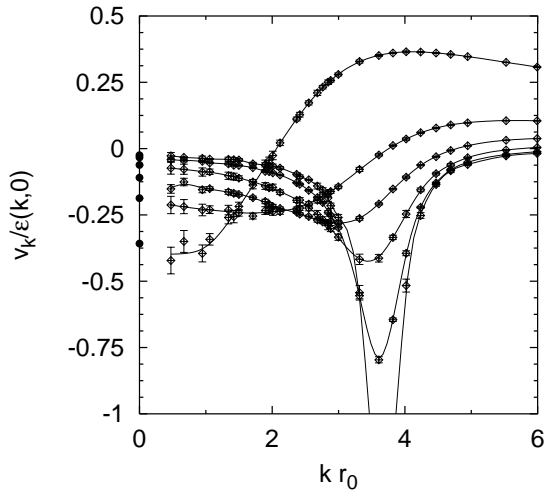}
\caption{ Effective interaction for $r_s=2,5,10,20,40$ and $60$
(open symbols, Monte Carlo data; lines, cubic spline interpolations).
Deeper minima correspond to lower densities. 
The solid dots at $k=0$ are the values of $1/\rho K_T$ from 
Tab.~ \ref{tab_ene}.} 
\label{effint}
\end{figure}

\begin{figure}[tb]
\includegraphics[width=67mm,angle=-90]{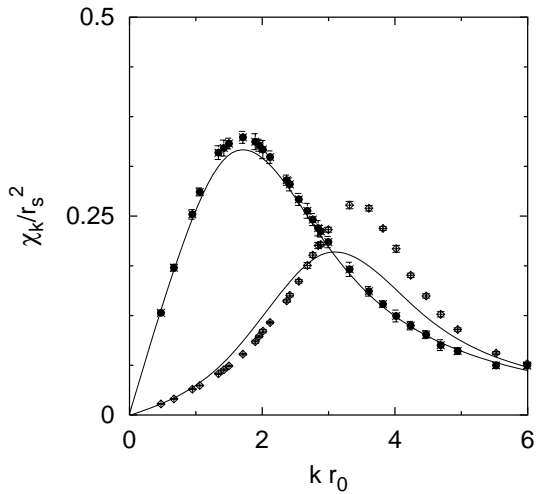}
\caption{The static response function $\chi(k)$
at $r_s=1$ (solid dots) and $r_s=10$ (open diamonds). The solid lines
are from Apaja {\it et al.}.\cite{saarela}  }
\label{Chi_cfr}
\end{figure}

In Fig.~(\ref{effint}) we report the static effective interaction
$v_k/\epsilon(k,0)$ where $v_k$ is the Coulomb interaction and 
$\epsilon(k,0)=1/[1 + v_k \chi(k,0)]$ is the static dielectric function.
At low $k$ the effective interaction is given the 
compressibility sum rule,
\begin{equation}
\lim_{k\rightarrow 0}\frac{v_k}{\epsilon(k,0)}=\frac{1}{\rho K_T},
\end{equation}
while in the short-wavelength limit it behaves like the Coulomb interaction.
The minimum of $v_k/\epsilon(k,0)$ deepens and shifts to larger $k$
upon increasing $r_s$.
We note that a negative dielectric function cannot be interpreted as 
a signal of instability of the bosonic fluid due to the presence of the 
rigid background. As in the case of the structural properties,
in the large coupling regime the Monte Carlo data for the effective 
potential show more pronounced features than the results of 
Apaya {\it et al.}.\cite{saarela} This is shown, in terms of the
static response function $\chi(k)$, in Fig.~(\ref{Chi_cfr}).

\subsection{Excitation spectrum}

The elementary excitations spectrum of the density fluctuation is
contained in the dynamic structure factor:
\begin{equation}
S(k,\omega)=\sum_{n}|\langle n|\rho_k|0\rangle|^2 
\delta(\omega-\omega_{n0}).
\end{equation}

We estimate the energy dispersion of the collective excitation
by fitting the imaginary time dependence of $F(k,\tau)$ with 
$F(k,\tau)=A(k)e^{-\omega_{1}(k)\tau}+
B(k)e^{-\omega_2(k)\tau}$. This amounts to represent the dynamical
structure factor $S(k,\omega)$ as the sum of two delta functions.
When a single mode has a dominating spectral weight, its dispersion
$\omega_1(k)$, is reproduced reasonably well,\cite{baroni} regardless of
the representation chosen for the remaining part of the spectrum 
(a delta function at $\omega_2(k)$ in this case).

Moreover, combining  our results for
$\chi(k)$ and $S(k)$ we obtain, by means of a sum-rules 
approach,\cite{Sum-rule,conti} a rigorous upper bound for the plasmon 
dispersion:
\begin{equation}\omega_k^{min} \le  
\frac{2\rho S(k)}{\chi(k)}.
\label{bound}
\end{equation}
At low $k$ a single mode exhausts the sum rule.
In this case, the upper bound in Eq.~(\ref{bound}) becomes an equality
and the strength of the excitation coincides with $S(k)$.
\begin{figure}[tb]
\includegraphics[width=67mm,angle=-90]{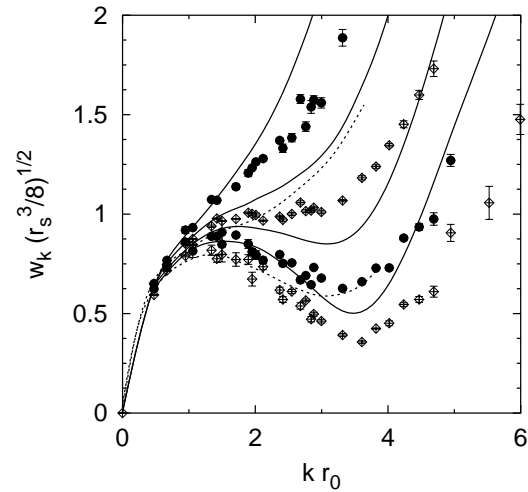}
\caption{The excitation spectra 
for $r_s=2,5,10,20$ (full circles and open diamonds) 
are compared with their respective 
upper-bound $\omega_k^{min}$ (solid lines).
Dashed curves corresponds to data from Ref.~\onlinecite{saarela}
for $r_s=5$ and $r_s=20$.
Curves with deeper minimum corresponds to lower densities.} 
\label{boundchi}\end{figure}

\begin{figure}[tb]
\includegraphics[width=67mm,angle=-90]{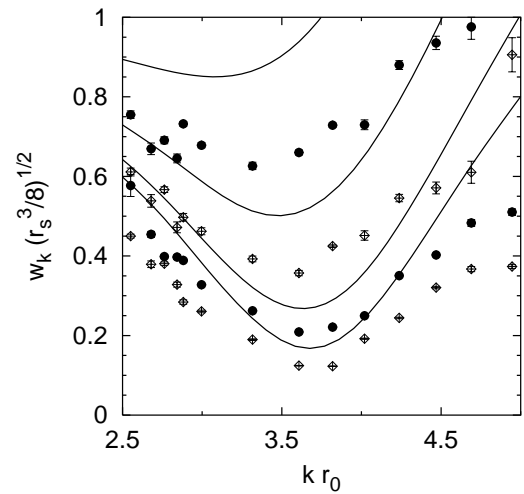}
\caption{Excitation spectrum near the rotonlike minimum 
for $r_s=10,20,40,60$. Full circles and open diamonds,
data from two-exponentials fit to $F(k,\tau)$; solid lines,
upper-bounds from Eq.~\ref{bound}.
}
\label{rotoni}\end{figure}

In Fig.~\ref{boundchi} 
we show our results for the excitation energies extracted directly 
from $F(k,\tau)$ and compare them with their corresponding
upper-bounds, at different densities. On increasing 
$r_s$ a roton--like mode, close to the first reciprocal lattice
vector of the Wigner crystal, develops and softens. The
evolution of this minimum as the crystallization transition is
approached is shown in more detail in Fig.~(\ref{rotoni}).

In conclusions, we have presented an extensive QMC study of
ground-state properties of 2D charged bosons. The present results
constitute a valuable benchmark for theoretical approaches, showing
their range of validity.

\end{document}